\documentclass[letter]{aa} 

\usepackage{natbib}
\bibpunct{(}{)}{;}{a}{}{,} 

\usepackage{graphicx}
\usepackage{mathtools, esdiff}
\newcommand{\dint}[1]{~\mathrm{d}#1}
\usepackage[varg]{txfonts}

\usepackage[print-unity-mantissa = false, range-units = single]{siunitx}
\DeclareSIUnit[]{\kb}{k_B}
\DeclareSIUnit[]{\yr}{yr}
\DeclareSIUnit[]{\Gyr}{\giga\yr}
\DeclareSIUnit{\bary}{m_u}
\DeclareSIUnit{\MJ}{M_J}
\DeclareSIUnit{\ME}{M_E}
\DeclareSIUnit{\RJ}{R_J}
\DeclareSIUnit{\kbperbary}{\kb \per \bary}
\DeclareSIUnit{\erg}{erg}
\DeclareSIUnit{\bar}{bar}

\usepackage{xspace}
\newcommand{\mesa}{\texttt{MESA}}
\newcommand{\Z}[1][\xspace]{heavy-element mass fraction\xspace}
\newcommand{\ZbulkT}[1][\xspace]{bulk \Z[1]}
\newcommand{\ZatmT}[1][\xspace]{atmospheric \Z[1]}

\usepackage[normalem]{ulem}

\newcommand{\Zbulk}{Z_\mathrm{bulk}}
\newcommand{\Zatm}{Z_\mathrm{atm}}
\newcommand{\Zcore}{Z_\mathrm{core}}
\newcommand{\mmid}{m_\mathrm{mid}}

\newcommand{\dSCool}{\Delta S_\mathrm{cool}}
\newcommand{\mRCB}{m_\mathrm{RCB}}
\newcommand{\ds}{\Delta s}
\newcommand{\dsComp}{\Delta s_\mathrm{comp}}

\begin{document} 

\title{Unraveling the origin of giant exoplanets}
\subtitle{Observational implications of convective mixing}
\author{H. Knierim\and R. Helled}
\institute{Department of Astrophysics, University of Zurich, Winterthurerstrasse 190, CH-8057 Zurich, Switzerland\\
\email{henrik.knierim@uzh.ch}
}
\date{Received 12 March 2025; Accepted 16 April 2025}
\abstract{
The connection between the atmospheric composition of giant planets and their origin remains elusive. 
In this study, we explore how convective mixing can link the primordial planetary state to its atmospheric composition. 
We simulate the long-term evolution of gas giants with masses between 0.3 and \SI{3}{\MJ}, considering various composition profiles and primordial entropies (assuming no entropy-mass dependence). 
Our results show that when convective mixing is considered, the atmospheric metallicity increases with time and that this time evolution encodes information about the primordial planetary structure. 
Additionally, the degree of compositional mixing affects the planetary radius, altering its evolution in a measurable way.
By applying mock observations, we demonstrate that combining radius and atmospheric composition can help to constrain the planetary formation history.
Young systems emerge as prime targets for such characterization, with lower-mass gas giants (approaching Saturn’s mass) being particularly susceptible to mixing-induced changes.
Our findings highlight convective mixing as a key mechanism for probing the primordial state of giant planets, offering new constraints on formation models and demonstrating that the conditions inside giant planets shortly after their formation are not necessarily erased over billions of years and can leave a lasting imprint on their evolution.}
\maketitle
\section{Introduction}
Unraveling the origin of gas giants is a fundamental goal of planetary science.
Naturally, the advent of accurate atmospheric measurements from space missions such as JWST \citep[][]{Gardner2006} and ARIEL \citep[][]{Tinetti2018} raises the question of how these data can advance this effort.
Planets form in disks with diverse chemical and physical conditions. It has been suggested that the planetary formation history leaves an imprint on their present-day composition \citep[e.g.,][]{Oeberg_2011, Schneider_2021a, Turrini_2021, Knierim_2022b}, although deciphering this imprint is not always straightforward \citep[e.g.,][]{Molliere_2022}.
Translating atmospheric abundance measurements into constraints on the bulk composition -- and, by extension, formation pathways -- remains a major challenge.
\par
Traditionally, atmospheric abundances have been interpreted as the planetary bulk composition 
(i.e., assuming that the planet is homogeneously mixed) or as the envelope's composition above a heavy-element core \citep[e.g.,][]{Sing_2024}.
Recent models have begun refining this framework; for instance, by integrating atmospheric and interior retrievals \citep{Acuna_2024, Wilkinson_2024}. However, the underlying assumption of a compositionally homogeneous planet or a distinct core-envelope structure remains largely unchanged.
In contrast, internal structure models of Jupiter and Saturn that fit the gravity data from the Juno \citep[e.g.,][]{Bolton2017} and Cassini \citep[e.g.,][]{Spilker_2019} missions suggest that both gas giants deviate from this simplified picture. Instead, they have inhomogeneous interiors and fuzzy cores \citep[see recent review by][and references therein]{HS2024}.
Building on these insights, \citet{Bloot_2023} developed an interior structure retrieval code with more intricate composition profiles.
Although an important step forward, this approach assumes a static composition, neglecting compositional changes through mixing.
To properly interpret atmospheric measurements of gas giant exoplanets, it is essential to model the mixing (of a given primordial composition profile) and determine the composition profiles inside giant planets for various conditions.\\ 
\indent In \citet{Knierim_2024} (hereafter KH24), we investigated the erosion of primordial composition gradients and showed that the survival of planetary cores depends on their shapes and primordial entropies. Moreover, we found that steep composition gradients with high metallicities and low primordial entropies (i.e., initially cold) can persist for gigayears, leading to atmospheric abundances that differ significantly from the planetary bulk composition. In contrast, low-metallicity profiles with high primordial entropies (i.e., initially hot) erode rapidly ($\sim\SI{1e7}{\yr}$), leading to the opposite trend of well-mixed planets, whereby the atmospheric composition is the same as the bulk composition.
Focusing on self-consistently reproducing present-day Jupiter and Saturn, \citet{Tejada_Arevalo_2025} and \citet{Sur_2025} independently confirmed these findings using their evolution code \texttt{APPLE} \citep{Sur_2024}.
Our previous study laid the theoretical foundation for understanding mixing processes in giant planets. In this work, we examine the observational implications, in particular the effect on the atmospheric abundances and the planetary radius. Moreover, we show how combining these two measurements can constrain the primordial planetary conditions, providing important insights for formation models.
This letter is structured as follows. Section \ref{sec:methods} outlines our methods. 
In Sect \ref{sec:results}, we investigate how convective mixing influences the atmospheric metallicity and radius evolution. 
Our discussion and conclusions are presented in Sects. \ref{sec:discussion} and \ref{sec:conclusion}, respectively.
%
\section{Methods}\label{sec:methods}
    The methods of this study are similar to the ones introduced in KH24. In short, we simulated the evolution of the planets using a modified version of the ``Modules for Experiments in Stellar Astrophysics'' code \citep[\mesa;][]{Paxton_2011, Paxton_2013, Paxton_2015, Paxton_2018, Paxton2019, Jermyn_2023}. Our modified version uses equations of state (EoSs) that are appropriate under planetary conditions \citep{Mueller_2020} and an improved treatment of convection and convective boundaries (KH24). For hydrogen and helium, we use the CD EoS \citep{Chabrier_2021}, while all elements heavier than these are represented by a 50-50 mixture of water (H$_2$O) and rock (SiO$_2$; \citealt{Mueller_2020}.
    All models utilize the semi-gray atmosphere model from \citet{Guillot2010}, assuming an equilibrium temperature of $T_\mathrm{eq} = \SI{500}{\K}$ (see Appendix \ref{sec:atm} for details).
    Unlike in KH24, we assume that semi-convective regions are stable, thereby damping the erosion of the core compared to the previous study (see Sect. \ref{sec:discussion} for details).
    Helium rain is not included in the models shown here due to its limited effect on the planetary evolution. A further discussion on helium rain is presented in Appendix \ref{sec:helium_rain}.
    \par
    We created the initial models for our numerical experiments with a proto-solar composition \citep{Lodders_2021}, before using the \texttt{relax\_initial\_entropy} and \texttt{relax\_initial\_composition} algorithms to create the desired entropy and composition profiles. 
    We note that the (specific) entropies we refer to throughout this study are the initial values for a protosolar composition before relaxing the composition profile (see Appendix \ref{sec:initial_models} for details).
    We varied these initial entropies from 8 to \SI{11}{\kbperbary}, ranging from cold to hot start scenarios.
    The composition profiles we relaxed are the ``core,'' ``extended,'' ``Helled 2023,'' and ``large core'' models from KH24. These profiles range from shallow, extended profiles to more traditional core-envelope structures (see also Appendix \ref{sec:initial_models}).
    Although these profiles are not meant to accurately represent any specific formation scenario, they qualitatively cover the expected behavior from formation studies \citep[e.g.,][and references therein]{Helled2023}.
    The planetary masses range from \SI{0.3}{\MJ} to \SI{3}{\MJ}, where \si{\MJ} is Jupiter's mass. We did not scale the composition profiles to the respective planetary mass. Instead, we simply added an envelope of protosolar composition on top of the composition profile. As a result, more massive planets have lower metallicities by construction.
    This setup is based on the assumption that most of the heavy elements in the planet are accreted during the early phases (phase-1, phase-2), followed by gas accretion that essentially determines the final planetary mass \cite{Helled2023}. 
    We evolved all models for \SI{1e3}{\yr} before initiating mixing, and then up to \SI{10}{\giga \yr}.
\section{Results}\label{sec:results}
\subsection{Atmospheric abundance evolution}\label{sec:Z_atm_vs_age}
\begin{figure*}
\centering
   \includegraphics[width=17cm]{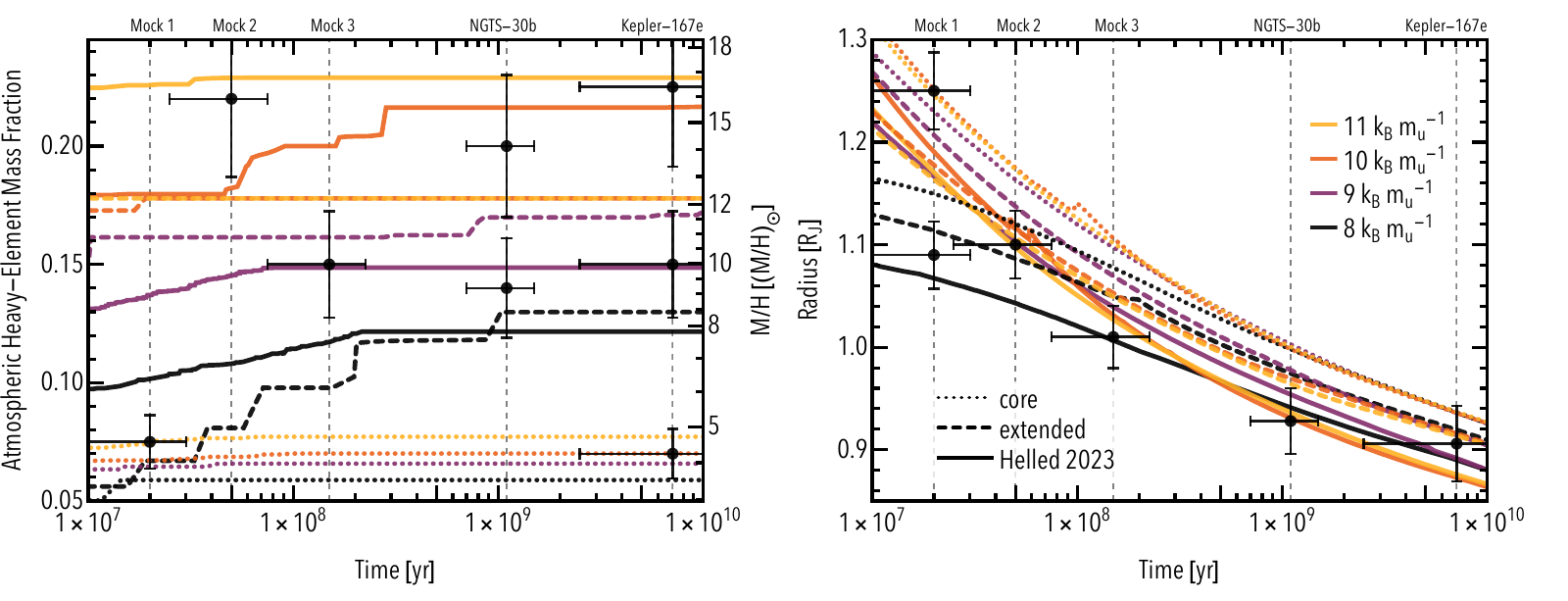}
     \caption{Left: Atmospheric heavy-element mass fraction vs. time for different primordial entropies and composition profiles (see legend in right plot). All models assume a \SI{1}{\MJ} planet.
     The data with errors are hypothetical atmospheric abundance measurements (see Sect. \ref{sec:mock_obs}).
     Right: Radius vs. time for the same models as in the left plot. The age and radius of NGTS-30b and Kepler-167e are measured (see Sect. \ref{sec:mock_obs}).
     }
     \label{fig:observable_evolution}
\end{figure*}

The left panel of Fig. \ref{fig:observable_evolution} shows the \ZatmT versus time for different primordial entropies of the Helled 2023, extended, and core models for a \SI{1}{\MJ} planet.
In all the simulations, the \ZatmT increases with time. Higher primordial entropies lead to higher \ZatmT values and faster mixing. 
Relative to their bulk metallicities of $\Zbulk = 0.25$ for the Helled 2023 model and $\Zbulk = 0.18$ for the extended model, the shallower composition profile of the extended model mixes more thoroughly.
Indeed, for a primordial entropy of \SI{11}{\kbperbary}, the extended model becomes fully mixed within \SI{1e7}{\yr}, whereas the Helled 2023 model retains a core.
The core model has the steepest and highest-Z composition profile and is therefore the most stable against convection. As a result, while the outer part of the gradient erodes, the inner core remains intact, leading to a plateau in the atmospheric metallicity plot. 
\par
The evolution of $\Zatm$ is dictated by three factors: the planetary cooling rate, the entropy profile, and the composition profile. Initially, the planet has an outer convective region and an inner radiative region (i.e., it is not adiabatic). As the planet cools, its outer convective region expands inward, thereby eroding the compositionally inhomogeneous interior.
This expansion is opposed by any negative $Z$ gradient and any positive entropy gradient.
Therefore, mixing is inefficient in planets with low-entropy (``cold'') interiors and steep composition gradients. 
Furthermore, because hotter planets cool faster, they also mix faster (for details on convective mixing, see KH24).
This interplay between the primordial entropy and the primordial composition profile leads to an increase in the planet's atmospheric metallicity over time.
Importantly, the primordial planetary structure dictates the evolution of its atmospheric abundances (see Appendix \ref{sec:analytic_details}).\\
\indent Therefore, if gas giants, or a subpopulation of them, form in a similar fashion (i.e., with similar composition profiles and entropies), their origin would be reflected in their atmospheric composition over time. On the other hand, the absence of such a pattern would suggest that gas giants originate with widely varying primordial structures.
Thus, if a predominant formation pathway exists, atmospheric abundance changes caused by convective mixing could provide a novel constraint on the planetary origin, particularly their primordial structure.
\subsection{Mixing influences the observed radius}\label{sec:radius_evolution}
The right panel of Fig. \ref{fig:observable_evolution} shows the planetary radius versus time for the simulations considered in Sect. \ref{sec:Z_atm_vs_age}.
Initially, hotter planets exhibit larger radii, but as they cool over billions of years, their sizes decrease, converging with the ones of their cooler counterparts.
Differences between primordial composition gradients, however, lead to more pronounced variations in radius.
While this result is well known, our models include an often-overlooked aspect: convective mixing. Mixing significantly alters the radius evolution in two ways. First, it redistributes material, which itself leads to different densities and material functions (e.g., opacities) throughout the planet. Second, it consumes energy, thereby reducing the planet's thermal energy content, and ultimately leading to smaller radii (see Appendix \ref{sec:radius_evolution_details} for a detailed comparison). These processes can change the radius beyond the measured accuracy, and therefore affect the inferred planetary composition (see Fig. \ref{fig:radius_comparison}). Therefore, convective mixing plays a crucial role in the determination of the radius evolution.
Overall, our results suggest that the primordial planetary state is not always erased with time (as in adiabatic models), and can leave an observational imprint. 
\subsection{Mock data}\label{sec:mock_obs}
Since both atmospheric metallicity and radius depend on the primordial planetary structure, combining these measurements can help infer the initial state. In this section, we illustrate this synergy by analyzing mock exoplanet data alongside available observations. For the mock data, we assumed relative errors of \qty{15}{\percent} for the \ZatmT, \qty{50}{\percent} for the age, and \qty{3}{\percent} for the radius, based on optimistic constraints from recent JWST studies \citep[e.g.,][]{Sing_2024,Welbanks_2024}.
For the observational data, we selected the exoplanets NGTS-30b \citep[\SI{0.96 \pm 0.06}{\MJ}, \SI{0.928 \pm 0.032}{\RJ}, \SI{1.1 \pm 0.4}{\Gyr};][]{Battley_2024} and Kepler-167e \citep[$1.01^{+0.16}_{-0.15}$ \si{\MJ}, \SI{0.91 \pm 0.04}{\RJ}, $7.1^{+4.4}_{-4.6}$ Gyr;][]{Chachan_2022}. These planets have masses comparable to Jupiter, equilibrium temperatures below \SI{1000}{\K} (i.e., are not inflated, see Sect. \ref{sec:discussion}), and radii that align with our simulations. 
For NGTS-30b and Kepler-167e, we considered two and three hypothetical atmospheric measurements, respectively, with uncertainties matching the ones of the mock data.
This combined dataset, shown in Fig. \ref{fig:observable_evolution}, forms the basis for the analysis presented below.

Planet ``Mock 1'' serves as our first example. With an estimated age of \SI{2 \pm 1e7}{\yr} and an atmospheric metallicity of $\Zatm = 0.075 \pm 0.011$, this young object aligns with core models above \SI{9}{\kbperbary} and the extended model at \SI{8}{\kbperbary}. These two models, however, predict notably different radii: $\sim\SI{1.25}{\RJ}$ for the core models above \SI{9}{\kbperbary} and $\sim\SI{1.12}{\RJ}$ for the \SI{8}{\kbperbary} extended model. This significant difference suggests that accurate radius measurements, as is shown in Fig. \ref{fig:observable_evolution}, could effectively distinguish between these competing formation pathways. 
In contrast, the planet ``Mock 2'' presents a more ambiguous case. Its radius of \SI{1.1 \pm 0.03}{\RJ} at an age of \SI{5 \pm 2.5 e7}{\yr} is consistent with most models, excluding only the \SI{8}{\kbperbary} Helled 2023 and larger than \SI{9}{\kbperbary} core configurations. A measured atmospheric metallicity of $\Zatm = 0.22 \pm 0.03$, however, would exclude all extended and core models, leaving only the Helled model with primordial entropies of \SI{10}{\kbperbary} or higher as possible options. 
Similarly, planet ``Mock 3'' demonstrates the power of combining multiple observables. With a radius of \SI{1.01 \pm 0.03}{\RJ} and an age of \SI{1.5 \pm 0.75e8}{\yr}, this exoplanet aligns with all Helled 2023 models. Including an atmospheric heavy-element abundance of $\Zatm = 0.15 \pm 0.02$ would rule out the 8, 10, and \SI{11}{\kbperbary} Helled 2023 models, setting limits on the primordial entropy.
\par
Turning to the observed data, the mass, radius, and age of NGTS-30b are consistent with all Helled 2023 models and 10 to \SI{11}{\kbperbary} extended models. However, the atmospheric metallicity predicted by the 8 and \SI{9}{\kbperbary} Helled 2023 models is $\Zatm \lesssim 0.15$. An accurate measurement of $\Zatm$, such as the ones indicated in Fig. \ref{fig:observable_evolution}, could exclude some of these scenarios.\\
\indent Finally, Kepler-167e is more challenging when it comes to discriminating between the different models. Its observed radius, mass, and age align with all the models considered in this study. Nonetheless, atmospheric abundance measurements could still discern between competing theories. Specifically, these measurements could distinguish the \SI{11}{\kbperbary} Helled 2023 model, the 8 and \SI{9}{\kbperbary} extended models, and the core models.\\
\indent These examples clearly show the importance of combining the measured mass, radius, and atmospheric metallicity with evolution models and the key role of convective mixing in exoplanetary characterization. 
In addition, combining these measurements can be used to exclude certain formation pathways or place upper or lower bounds on the system’s primordial entropy.
Younger planetary systems are particularly promising, as both their radii and atmospheric metallicities tend to exhibit the largest differences between various primordial structures.
For example, different primordial composition gradients such as the \SI{8}{\kbperbary} extended model and the \SI{10}{\kbperbary} core model differ by more than \SI{0.1}{\RJ}, making them readily distinguishable with current observational capabilities.
Detecting and characterizing young exoplanets remains challenging, but recent discoveries \citep[e.g.,][]{Barber_2024} show that these systems are observable, with the number of detected young exoplanets steadily increasing.
As such planets evolve, their radii tend to converge within the observational uncertainties assumed in this study. Even in such cases, atmospheric metallicity measurements remain valuable, as they can be used to exclude certain primordial structures.
The effectiveness of these constraints, however, depends on both the quality of the observational data and the region of parameter space being probed. Some areas of the radius-metallicity-age space are inherently more degenerate, making it more challenging to distinguish between formation scenarios (see Sect. \ref{sec:discussion}).
Also, to fully interpret any individual object, a broader range of primordial entropies and composition profiles would need to be explored.
%
\subsection{Radius evolution for different planetary masses}\label{sec:radius_evolution_w_mass}
\begin{figure*}
\centering
   \includegraphics[width=15cm]{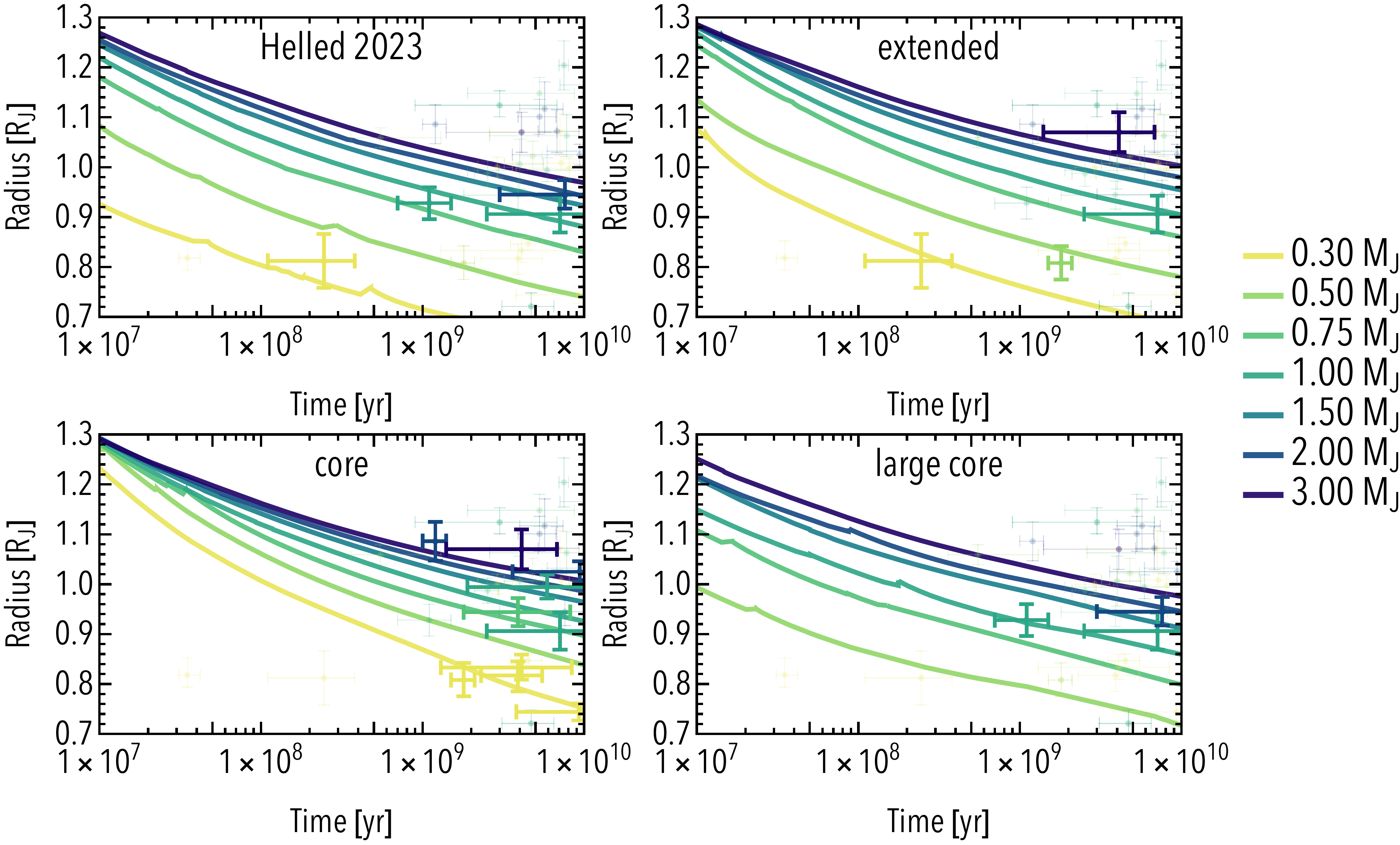}
     \caption{Radius evolution of different planetary masses for the Helled 2023 model (\textit{top left}), extended model (\textit{top right}), core model (\textit{bottom left}), and large core model (\textit{bottom right}). The large core model for $\SI{0.3}{\MJ}$ is below the shown range. The exoplanet data were obtained from the PlanetS catalog \citep{Otegi_2020, Parc_2024}. Data points consistent with the evolutionary tracks are highlighted with thick lines, while ones that deviate are shown with thin lines.}
     \label{fig:radius_evolution_for_different_masses}
\end{figure*}

Figure \ref{fig:radius_evolution_for_different_masses} shows the radius evolution of all models for different planetary masses.
Beyond the trends already discussed in Sect. \ref{sec:radius_evolution}, we observe that planetary radii converge with increasing planetary mass.
This behavior is primarily driven by the well-known mass-independent radius of hydrogen-helium (H-He) -dominated planets \citep{Zapolsky_1969}.
Within our model framework, more massive planets possess larger H-He envelopes, resulting in lower bulk metallicities and radii that increasingly approach the ones expected from a pure H-He planet.
A more subtle effect is the reduced mixing efficiency in more massive planets at a given primordial entropy (KH24). As a result, more massive planets develop similar metallicity profiles (in absolute mass coordinates), leading to weaker mixing-induced variations in radius.
On the other hand, the lower the planetary mass, the more the planetary radii differ as their bulk metallicities and their mixing properties increasingly deviate from each other.
Consequently, under the assumption that more massive planets have larger H-He envelopes, the interior structures of lower-mass planets are less degenerate than the ones of higher-mass planets, allowing for stronger constraints on their primordial conditions. We note, however, that this is a strong assumption that may not always hold. It is clear that further investigations in this direction are required. \\
\indent In order to compare our models with observations, we included all exoplanets from the PlanetS catalog \citep{Otegi_2020, Parc_2024} that fall within our planetary mass range and have equilibrium temperatures below \SI{1000}{\K} (see Sect. \ref{sec:discussion}). Exoplanets consistent with a given model’s evolutionary track (in terms of mass, age, and radius) are highlighted with thick lines.
We find that although different models reproduce subsets of the observed data to varying degrees, none of them can explain all exoplanets in the sample.
This is expected, given the significant spread in observed radii even for a fixed planetary mass. For example, at \SI{1}{\MJ}, the radii of observed exoplanets span more than \SI{0.5}{\RJ}, implying that there is no single formation pathway that can explain the entire population. Instead, a diversity of primordial structures is required.
The current dataset remains relatively sparse, with only seven exoplanets in the PlanetS catalog that match \SI{1}{\MJ} and $T_\mathrm{eq}<\SI{1000}{\K}$.
As observational surveys expand, a clearer picture may emerge, potentially revealing new trends in planetary structure and evolution.
\section{Discussion}\label{sec:discussion}
    The results presented here are subject to two key and distinct sources of uncertainty: (1) simplifications in the physical modeling, and (2) ambiguities in the initial conditions.
    On the modeling side, uncertainties arise from a simplified treatment of convection (mixing length theory), the atmosphere model, and neglected effects such as rotation \citep{Fuentes_2023}. In addition, the material properties, specifically the opacities and EoSs, remain poorly constrained \citep[e.g.,][]{Cozza_2025} and can influence a planet's evolution \citep[][]{Mueller_2024, Howard2025}.
    Also, the planetary metallicity was represented by an ideal mixture of water and rock, while in reality various heavy elements can exist in the interiors of giant planets (and at different depths), influencing the mixing efficiency, and therefore the long-term evolution.
    \par
\begin{figure}
    \centering
     \resizebox{7.5 cm}{!}{\includegraphics{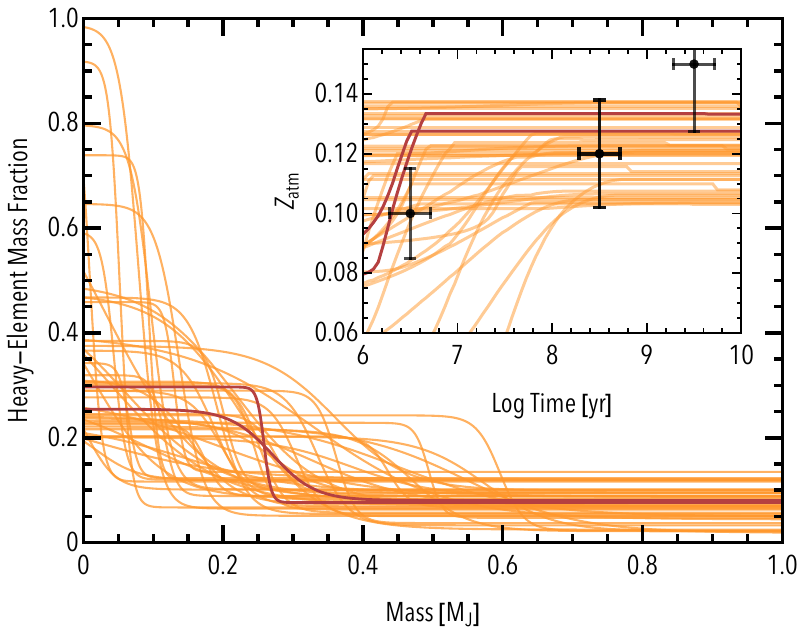}}
      \caption{Primordial heavy-element mass fraction vs.~mass. The panel in the top right corner shows the atmospheric metallicity vs.~time for these curves, created with the analytic model from KH24 (for details, see Appendix \ref{sec:degeneracy_details}).
      The orange curves are consistent with $\Zatm=0.12 \pm 0.02$ at $\SI{3.2 \pm 1.6e8}{\yr}$ (bold mock data point). The red curves are in addition consistent with with $\Zatm=0.1 \pm 0.02$ at $\SI{3.2 \pm 1.6e6}{\yr}$ and $\Zatm=0.15 \pm 0.02$ at $\SI{3.2 \pm 1.6e9}{\yr}$ (thin mock data points).}
      \label{fig:degeneracy}
    \end{figure}

    Here, we have investigated four distinct primordial composition profiles and four different primordial entropy profiles. Specifically the primordial entropy, however, remains unknown and could also depend on the planetary mass.
    Therefore, we have also considered cases in which more massive planets have the same primordial core but higher primordial entropies than their less massive counterparts (see Appendix \ref{sec:mass_trend}).
    While our models only represent a small subset of all the physically plausible primordial structures, they  illustrate the additional information gained by considering the atmospheric composition.  
    To illustrate this, Fig. \ref{fig:degeneracy} presents a range of distinct primordial structures that are all consistent with mock observations.
    \par
    In addition, the ability to constrain primordial structures remains limited by the lack of observed non-inflated gas giants. The highest-quality observations typically come from close-in inflated gas giants. As long as the inflation mechanism and its magnitude remain unknown, it is not possible to infer their composition and internal structure accurately.
    \par
    Furthermore, observations retrieve specific molecular abundances in the upper atmosphere (millibar to bar) while our inferred atmospheric metallicity, $\Zatm$, represents the heavy-element mass fraction in the outer convective envelope (\qtyrange{1}{1e3}{\bar}). In addition, we represent the heavy elements with an idealized mixture of water and rock. As a result, directly translating observed molecular abundances into an envelope metallicity is not straightforward and requires further investigation. 
    \par
    Finally, a better understanding of planet formation could provide more stringent constraints on the planetary initial compositional and thermal structure, thereby excluding some evolutionary models that, while consistent with present-day observations, are unlikely to arise from realistic formation processes.
\section{Conclusions}\label{sec:conclusion}
    We have explored the observational consequences of convective mixing. We have showed that accounting for convective mixing can break part of the degeneracy between hot and cold start scenarios, allowing one to constrain the primordial state of a planet with current-day observational capabilities. Our key conclusions are that:
    \begin{enumerate}
        \item The atmospheric metallicity increases with time. The shape of the $\Zatm(t)$ curve is determined by the planet's primordial entropy and composition profile.
        \item Mixing alters the planetary radius by redistributing material.
        \item Combining mass, radius, and atmospheric metallicity can break degeneracies, thereby constraining the primordial state of the planet, particularly in young planets. 
        \item Low-mass gas giants, for which the primordial composition profile represents a substantial fraction of the total mass, are prime candidates for having their primordial states retrieved. The higher the planetary mass (the higher the H-He mass fraction), the harder it is to discriminate among different primordial states' formation histories. 
    \end{enumerate}
    Overall, our results indicate that the primordial planetary conditions are not always erased and can leave an observable imprint.  
        Therefore, convective mixing in giant planets can provide an important link between the primordial planetary state and present-day observables, offering a new avenue to constrain their formation history. As we enter an era of unprecedented observational precision with JWST and soon ARIEL, we are closer than ever to delivering on their promise: unraveling the origin of giant exoplanets. 
\begin{acknowledgements}
    This work has been carried out within the framework of the National Centre of Competence in Research PlanetS supported by the Swiss National Science Foundation under grants \texttt{51NF40\_182901} and \texttt{51NF40\_205606}.
\end{acknowledgements}

\bibliographystyle{aa} 
\bibliography{main} 
\begin{appendix} 
\section{Atmospheric boundary conditions} \label{sec:atm}
We adopt the default implementation of the semi-gray, globally averaged atmosphere model of \citet{Guillot2010}, as introduced in \citet{Paxton_2013}. Specifically, we use \mesa’s \texttt{atm\_irradiated\_opacity = ‘iterated’} option, in which the thermal opacity is iteratively adjusted to match the temperature and pressure at the base of the atmosphere. Below the atmosphere, the opacities are provided by \mesa’s opacity module and depend on the local thermodynamic conditions, including composition. The visible opacity is fixed and taken from \citet{Guillot2010}, who validated this choice by comparing to more detailed atmospheric models.
While a fully self-consistent treatment of atmospheric boundary conditions would be ideal, we emphasize that the mixing inside the planet is rather insensitive to the specific choice of atmospheric conditions (see KH24).
\section{Helium rain}\label{sec:helium_rain}
\begin{figure*}[h!]
\centering
   \includegraphics[width=17cm]{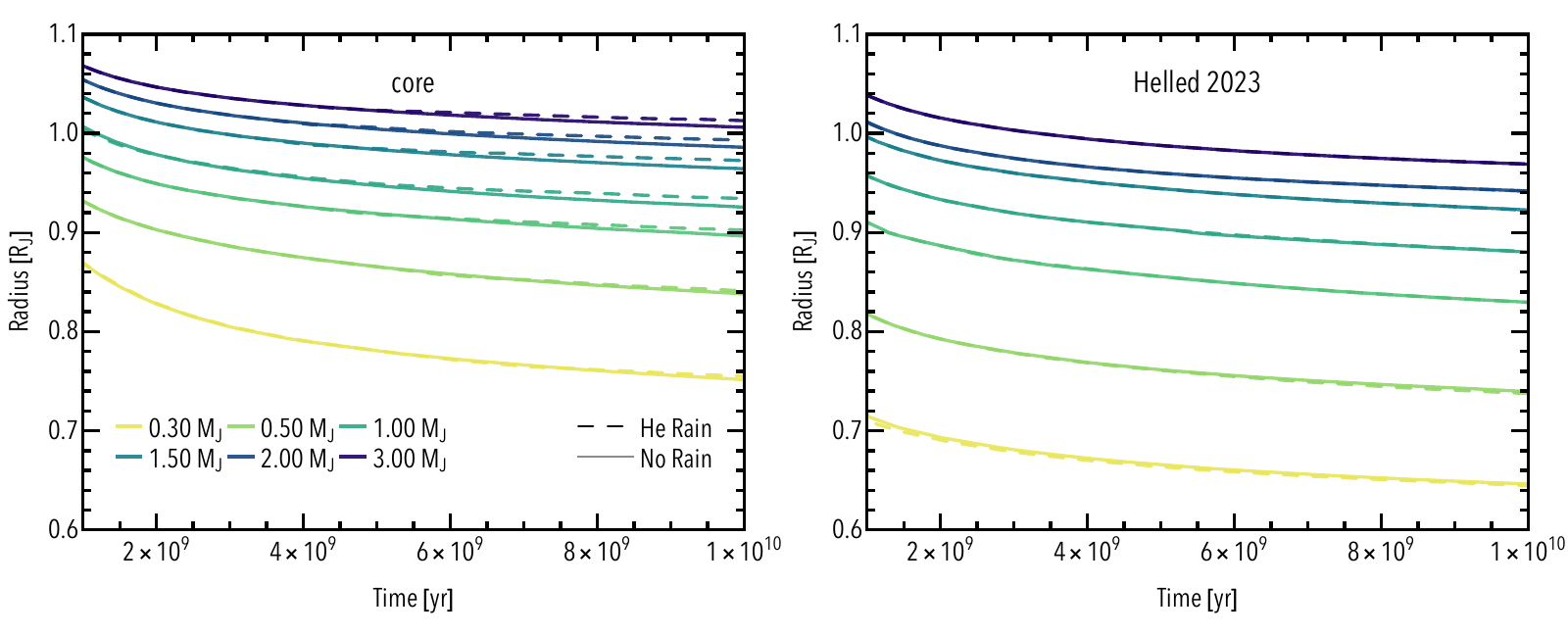}
     \caption{Radius evolution of different planetary masses with and without helium rain for the core model (\textit{left}) and the Helled 2023 model (\textit{right}).}
  \label{fig:helium_rain}
\end{figure*}

Helium rain refers to the process by which helium becomes immiscible in hydrogen under given pressure and temperature conditions, leading to phase separation between hydrogen and helium, and the formation of helium droplets that settle toward deeper layers \citep{Stevenson_1977b}.
This phase separation releases gravitational energy, which can delay the planetary cooling \citep{2003Icar..164..228F, Howard_2024}.
Observational evidence suggests that helium rain occurs in both Jupiter and Saturn \citep[see][and references therein]{HS2024}, but the exact conditions for helium immiscibility remain uncertain. Several phase diagrams which differ in the extent and location of the immiscibility region have been proposed \citep[e.g.,][]{Morales2009, lorenzen2011, Schoettler2018}.
As a result, the impact of helium rain on the thermal and structural evolution of giant exoplanets remains unclear.
To explore the effect of helium rain on our results, we conducted a series of simulations using the phase diagram of \citet{Schoettler2018}, assuming instantaneous removal of immiscible helium. Figure~\ref{fig:helium_rain} compares the radius evolution of models with and without helium rain.
We find that across the range of models considered, helium rain has only a small impact on the planetary evolution. This is because of two main reasons. First, the strong stellar irradiation delays the planetary cooling and therefore the onset of helium immiscibility.  Second, convective mixing enriches the envelope with heavy elements, which reduces the helium concentration, and thus the amount available to separate once immiscibility begins.
As a result, models with strongly enriched envelopes—such as Helled 2023 and extended, where $\Zatm$ exceeds \qty{10}{\percent}—experience little to no helium rain. In contrast, models with more stable composition gradients, such as core and large core, retain lower envelope metallicities and do show a small increase in radius after several gigayears. Still, the effect remains small.
In all cases, neither the increase in planetary radius nor the change in atmospheric metallicity caused by helium rain are currently observable. Our results imply that helium rain plays a limited role in the evolution of hot and warm Jupiters in comparison to cold giant planets. 
\section{Initial model details}\label{sec:initial_models}
\begin{figure*}
\centering
   \includegraphics[width=17cm]{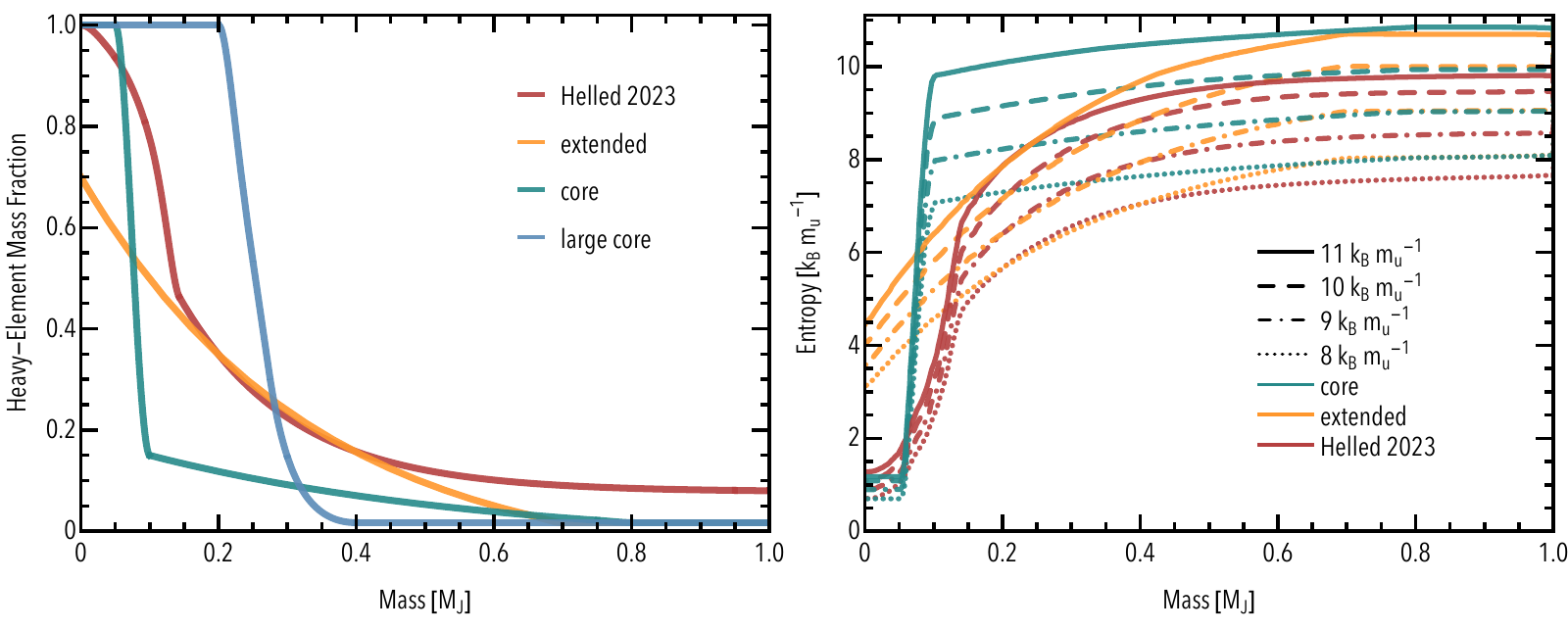}
     \caption{Left: Initial metallicity profiles of the four models investigated in this study. Right: Entropy profiles of models Helled 2023, extended, and core investigated in this study. The entropy labels in the legend indicate the specific (homogeneous) entropy of the proto-solar composition prior to relaxing the composition gradient.}
  \label{fig:initial_conditions}
\end{figure*}

Figure \ref{fig:initial_conditions} (left) shows the heavy-element mass fraction as a function of mass for the four models investigated in this study. 
These models are identical to those used in KH24 (see Fig. 3 in KH24).
Furthermore, Fig. \ref{fig:initial_conditions} (right) shows the initial entropy profiles for the numerical experiments presented in Fig.~\ref{fig:observable_evolution}. These profiles are not isentropic but exhibit a nearly linear increase in specific entropy with decreasing metallicity, reflecting the imposed compositional gradient.
To isolate the effects of thermal structure from those of composition, and to enable comparisons with previously published isentropic models, we define a reference entropy value based on an equivalent homogeneous planet with protosolar composition. Specifically, when we refer to an initial entropy of, for example, \SI{8}{\kbperbary}, we mean that the planet would have this entropy if it were fully convective and of protosolar composition.
In this sense, the quoted entropies can be interpreted as protosolar-equivalent entropies, providing a consistent benchmark across different interior structures.
\section{Analytic details}\label{sec:analytic_details}
Let $Z(m)$ be the primordial composition profile. As the outer convective region eats its way into the planet, the atmospheric metallicity increases to
\begin{align}
    \Zatm = \frac{1}{M-\mRCB}\int\limits_{\mRCB}^{M} Z(m) \dint{m},
\end{align}
where $M$ is the planetary mass and $\mRCB$ is the mass coordinate of the boundary between the inner radiative and the outer convective region.
As we showed in KH24, this boundary is given by
\begin{equation}
    \frac{\dSCool}{M-\mRCB} = \ds(\mRCB) + \dsComp(\mRCB),
\end{equation}
where $\dSCool$ is the (total) entropy loss, $\ds$ is the (specific) entropy difference between the core and the envelope, and $\dsComp$ is the same difference but for the composition entropy.
We define the composition entropy as the integral over the composition gradient multiplied by a response function that converts the gradient to an entropy.
Using the $X$, $Y$, and $Z$ notation for the mass fraction of hydrogen, helium, and everything else, respectively, as well as assuming a fixed ratio between $X$ and $Y$ (e.g., protosolar), the composition entropy simplifies to
\begin{align}
    s_\mathrm{comp} = \int\limits_{0}^{M} \diffp*{s}{Z}{P, \rho}\diff{Z}{m}\dint{m},
\end{align}
where $P$ and $\rho$ are the pressure and density, respectively. Thus, the function form of $\dsComp$ is directly influenced by the composition gradient $\dint{Z}/\mathrm{d}m$.
Over gigayears, planets cool to approximately the same entropy (i.e., most of the cooling happens early). Hence, in the absence of any additional mechanism that changes the primordial shape of $\ds$ or reduced the cooling rate $\dSCool$, the evolution of the atmospheric metallicity is largely governed by the primordial entropy and the primordial composition gradient (for more details, see KH24).
\section{Radius evolution details}\label{sec:radius_evolution_details}
\begin{figure}
  \resizebox{\hsize}{!}{\includegraphics{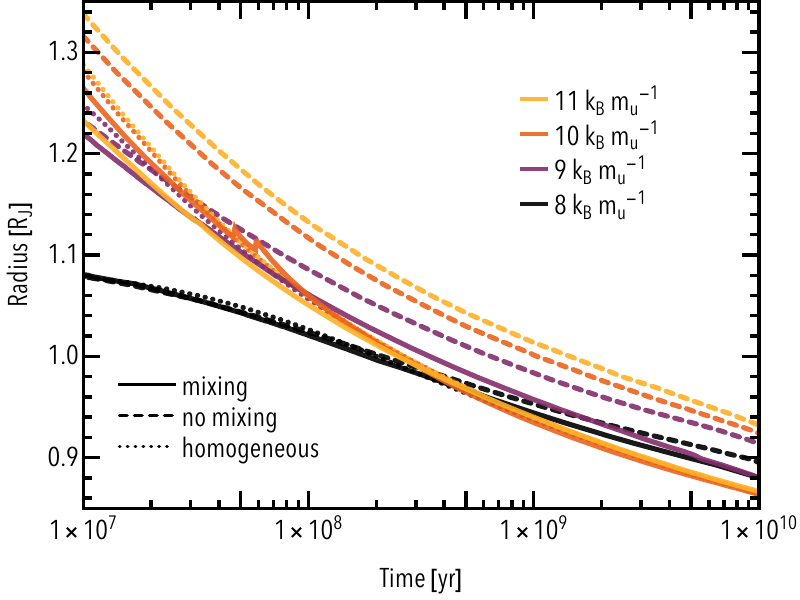}}
  \caption{Radius evolution of a \SI{1}{\MJ} Helled 2023 model for different entropies. The solid lines indicate models with convective mixing.  Dashed lines show the same models without mixing, while dotted lines represent planets with the same bulk metallicity, but with the heavy elements being homogeneously distributed.}
  \label{fig:radius_comparison}
\end{figure}

Figure \ref{fig:radius_comparison} shows the radius evolution of a \SI{1}{\MJ} Helled 2023 model under three different scenarios: (i) with convective mixing (our default case), (ii) without mixing, and (iii) with the same bulk metallicity but with the heavy elements being homogeneously distributed. 
Planets with a homogeneous distribution of heavy elements converge to the same radius after a few gigayears as their initial conditions are erased.
On the other hand, when mixing is suppressed, composition gradients inhibit convection, reducing the efficiency of the planetary cooling. This leads to a larger radius compared to the other models. 
 Notably, for the \SI{11}{\kbperbary} case, neglecting mixing results in an overestimate of the radius by about \SI{0.08}{\RJ}, which lies within current observational uncertainties. This highlights the importance of accounting for convective mixing when interpreting observational data. 
\section{Uncertainty in primordial entropy and its mass dependence}\label{sec:mass_trend}
In Sect. \ref{sec:radius_evolution_w_mass}, we assumed a constant primordial entropy of \SI{9}{\kbperbary} across all planetary masses.
In reality, however, the specific entropy is expected to vary with mass-particularly during runaway gas accretion, where shock heating establishes a positive entropy gradient in more massive envelopes \citep[e.g.,][]{Cumming2018}.
However, the entropy gradient built up during runaway gas accretion may not affect the mixing of primordial composition gradients significantly, as mixing begins only once the outer envelope has cooled sufficiently (KH24).
\par 

\begin{figure*}[h!]
\centering
   \includegraphics[width=17cm]{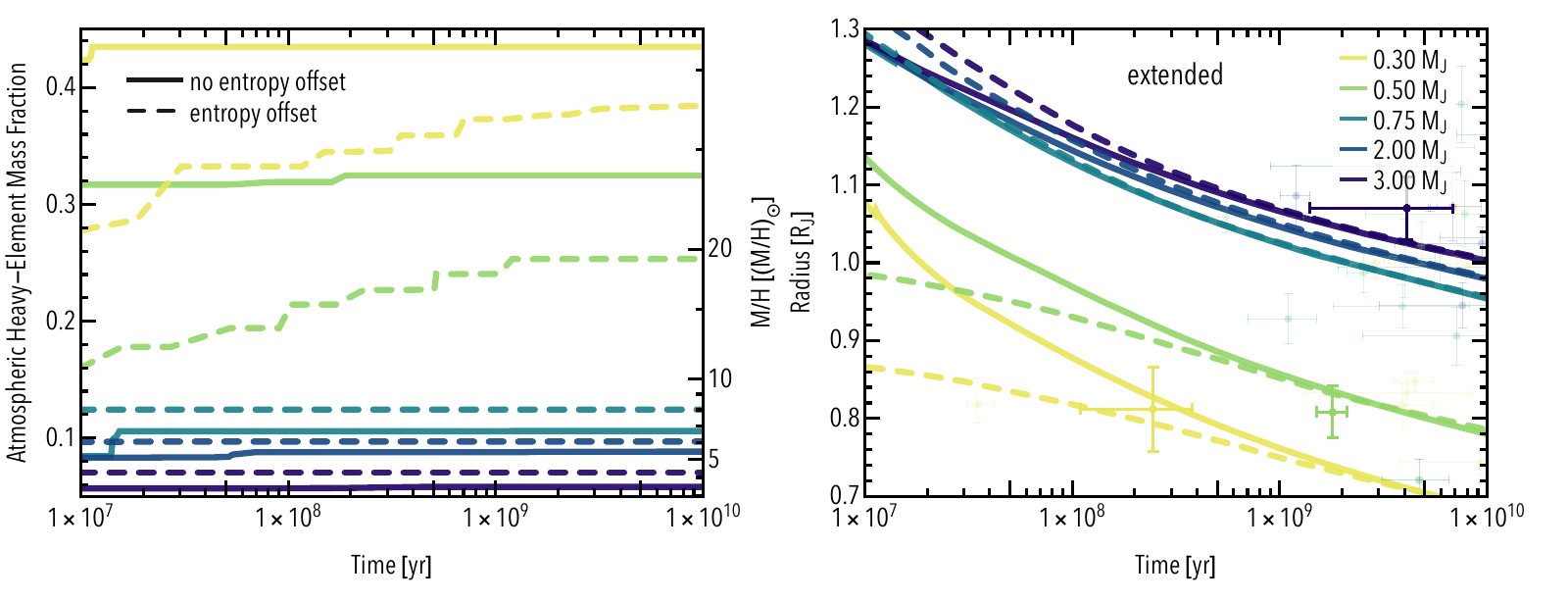}
     \caption{Left: Atmospheric metallicity over time for the “extended” composition model at different planetary masses. Solid curves correspond to the default entropy of \SI{9}{\kbperbary}; dashed curves indicate models with adjusted entropies. Right: Radius evolution for the same set of simulations. See text for further details.
     }
  \label{fig:entropy_scaling_mass}
\end{figure*}

The composition gradients in the deep interior correspond to the earlier formation phases and strongly depend on the accretion rates and the local formation conditions. 
It is yet to be determined whether the central entropy correlates with planetary mass.
To explore the impact of a potential correlation between core entropy and planetary mass, we repeated the simulations of model “extended” at various masses, adjusting the entropy relative to the \SI{9}{\kbperbary} baseline: by $-\SI{1}{\kbperbary}$ for \SI{0.3}{\MJ} and \SI{0.5}{\MJ}, by $+\SI{1}{\kbperbary}$ for \SI{1.5}{\MJ} and \SI{2.0}{\MJ}, and by $+\SI{2}{\kbperbary}$ for \SI{3.0}{\MJ}.
Figure \ref{fig:entropy_scaling_mass} compares these simulations to the  models used in the main text.
We find that varying the entropy alters the mixing efficiency and, consequently, the atmospheric metallicity. This effect is most pronounced in the low-mass models, consistent with our findings in Sect. \ref{sec:results}. In contrast, the radius evolution is rather insensitive to the entropy shift, with noticeable differences only for very cool, low-mass planets. 
These results demonstrate that while the exact outcome depends on the specific model assumptions, the overall trends remains. Therefore, it is clear that the consideration of convective mixing has an added value.
\section{Details of the degeneracy plot}\label{sec:degeneracy_details}
To compute all the curves shown in Fig. \ref{fig:degeneracy} in a reasonable time, we employed the analytic mixing model described in Appendix \ref{sec:analytic_details}. To obtain the $\ds$ and $\dsComp$, we assumed $\ds(m) = \alpha_s(s,Z) (Z(m)-Z(0))$ and $\dsComp(m) = \alpha_\mathrm{comp}(s,Z) (Z(m)-Z(0))$, where the response functions $\alpha_s$ and $\alpha_\mathrm{comp}$ were obtained by interpolating \mesa~simulations.
Similarly, we obtained the cooling rate $\dSCool$ by interpolating \mesa~simulations.
The metallicity curves were generated with a generalized logistic function,
\begin{align}
    Z(m) = \Zcore - \frac{\Zcore-\Zatm}{1+\exp\left[\alpha (m - \mmid)\right]},
\end{align}
where $\Zcore$ and $\Zatm$ are the heavy-element mass fractions in the limits of $m \to -\infty$ and $m \to \infty$, respectively. The steepness of the transition between $\Zcore$ and $\Zatm$ is quantified with $\alpha$, the location of the transition by the midpoint $\mmid$. For sufficiently steep values of $\alpha$, $\Zcore$ and $\Zatm$ can be thought of as the heavy-element mass fraction at the core and in the atmosphere.
To generate a large sample of initial models, we drew $\Zcore$ from a uniform distribution between 0 and 1, $\Zatm$ from a uniform distribution between 0 and $\Zcore$, $\mmid$ from a uniform distribution between 0 and \SI{1}{\MJ}, and $\alpha$ from a log-uniform distribution between 1 and 200.
\end{appendix}
\end{document}